\documentclass{article}

\usepackage{arxiv}

\usepackage[utf8]{inputenc} % allow utf-8 input
\usepackage[T1]{fontenc}    % use 8-bit T1 fonts
\usepackage{hyperref}       % hyperlinks
\usepackage{url}            % simple URL typesetting
\usepackage{booktabs}       % professional-quality tables
\usepackage{amsfonts}       % blackboard math symbols
\usepackage{nicefrac}       % compact symbols for 1/2, etc.
\usepackage{microtype}      % microtypography
\usepackage{lipsum}
\usepackage{graphicx}
\usepackage{amsmath}
\graphicspath{ {./images/} }

\begin{document}

\title{Differentiable model-based adaptive optics with transmitted and reflected light}

\author{
 Ivan Vishniakou$^*$ \\
  Center of Advanced European\\ Studies and Research (caesar)\\
  53175 Bonn, Germany
  %% examples of more authors
   \And
 Johannes D. Seelig\thanks{Corresponding authors: ivan.vishniakou@caesar.de, johannes.seelig@caesar.de} \\
  Center of Advanced European\\ Studies and Research (caesar)\\
  53175 Bonn, Germany\\
}

\maketitle
%%%%%%%%%%%%%%%%%%% abstract %%%%%%%%%%%%%%%%

\begin{abstract}
%This format begins with an introductory paragraph (not abstract) of 150 words maximum, summarizing the background, rationale, main results and implications.

%background
Aberrations limit optical systems in many situations, for example when imaging in biological tissue.
Machine learning offers novel ways to improve imaging under such conditions by learning inverse models of aberrations.
Learning requires datasets that cover a wide range of possible aberrations, which however becomes limiting for more strongly scattering samples, and does not take advantage of prior information about the imaging process.

Here, we show that combining model-based adaptive optics with the optimization techniques of machine learning frameworks can find aberration corrections with a small number of measurements. 
Corrections are determined in a transmission configuration through a single aberrating layer and in a reflection configuration through two different layers at the same time. Additionally, corrections are not limited by a predetermined model of aberrations (such as combinations of Zernike modes).
Focusing in transmission can be achieved based only on reflected light, compatible with an epidetection imaging configuration.
\end{abstract}

%%%%%%%%%%%%%%%%%%%%%%%%%%  body  %%%%%%%%%%%%%%%%%%%%%%%%%%
\section*{Introduction}
Machine learning offers novel approaches to correct for aberrations encountered when imaging though scattering materials, 
\cite{kerr2008imaging, rodriguez2018adaptive, rotter2017light, yoon2020deep}
from astronomy \cite{angel1990adaptive, paine2018machine, swanson2018wavefront, andersen2019neural} to microscopy with transmitted (for example \cite{jin2018machine, hu2019learning, cheng2019artificial}) and reflected light \cite{vishniakou2020wavefront}.
%problem area
To find aberration corrections in these situations, machine learning typically relies on large synthetic datasets. 
Large datasets are required, first, because the many parameters of deep neural networks need to be adjusted to  work under a wide range of conditions and all these conditions need to be covered in the training data.
Secondly, machine learning models are typically agnostic about the underlying image generating process. Therefore, even \textit{a priori} known information, for example the transformations inside the optical system, needs to be learned from data. 

%2. previous and current research contributions

%Machine learning approaches have been introduced in adaptive optics in astronomy and in adaptive optics for imaging in transmission. More recently, such approaches have also been generalized to imaging with reflected light, an approach related to multiconjugate optics where multiple layers of scattering are corrected. 
In practice, training datasets are often based on
%For good model performance in actual imaging situations, these datasets, which are generated in simulations, need to accurately cover the full range of potentially encountered aberrations. 
%A close resemblance between synthetic training data and experimentally observed aberrations is important since the trained model will likely not generalize well to situations that were not covered in the training data.
combinations of Zernike polynomials \cite{angel1990adaptive, paine2018machine, swanson2018wavefront, andersen2019neural, jin2018machine, hu2019learning, cheng2019artificial, vishniakou2020wavefront} which might however not accurately capture all aspects of experimentally encountered aberrations. Additionally, for more strongly scattering samples, which require increasingly higher orders of Zernike modes, covering all potential scattering situations by sampling a sufficient number of different mode combinations eventually results in very large datasets. This is in particular the case if aberrations in multiple layers are combined, for example when using reflected light in an epidetection configuration \cite{vishniakou2020wavefront}. 

While finding inverse models through such data driven strategies is well suited for situations where the underlying physical model is undetermined, the image formation process in an optical system is typically at least partially known. This is the basis of model-based adaptive optics, where optical systems modeling is combined with optimization to find an unknown phase aberration \cite{gonsalves2014perspectives, song2010model, yang2015model, antonello2015modal}. 
%Therefore, not taking advantage of such prior information and instead learning it anew from data can be considered inefficient.
Similar situations where models based on a well known underlying physical process are learned from data are also encountered in other imaging modalities \cite{ongie2020deep}, and more broadly many areas of engineering and physics %(for example \cite{loper2014opendr, li2018differentiable, degrave2019differentiable, giftthaler2017automatic, de2018end, heiden2019interactive, schenck2018spnets, vilsmeier2019transfer, heiden2019physics, kellman2019data, wang2020phase, bostan2020deep, zhou2020diffraction}). 
(for example \cite{loper2014opendr, li2018differentiable, degrave2019differentiable, heiden2019physics, kellman2019data, wang2020phase, bostan2020deep, zhou2020diffraction}). 
To take advantage of such prior information, methods have been developed that combine physical process models with machine learning optimization. 

For such optimization, first a model is described as a differentiable function mapping input to output. Since the function is differentiable, one can take advantage of automatic differentiation, which is more accurate and computationally efficient than finite differences \cite{baydin2017automatic, margossian2019review, lu2019deepxde}, and is one of the cornerstones of machine learning frameworks such as Tensorflow. Automatic differentiation is used in these frameworks to compute gradients for optimization of a loss function with respect to parameters of interest. The loss function compares model output to a target output and the discrepancy is minimized by adjusting model parameters  
(\cite{loper2014opendr, li2018differentiable, degrave2019differentiable, heiden2019physics, kellman2019data, wang2020phase, bostan2020deep, zhou2020diffraction}). 
%This approach has been implemented in different domains, examples include image rendering \cite{loper2014opendr, li2018differentiable}, rigid body dynamics and robotics \cite{degrave2019differentiable, giftthaler2017automatic, de2018end, heiden2019interactive}, fluid dynamics \cite{schenck2018spnets}, beamlines \cite{vilsmeier2019transfer}, LIDAR \cite{heiden2019physics}, as well as ptychography \cite{kellman2019data}.  

%A feed-forward artificial neural network as a universal approximator is able to represent \cite{cybenko1989approximation} a wide variety of functions, and if a sought mapping exists and the training process liske stochastic gradient descent is lucky to converge, a model approximating the sought unknown function can be acquired. 

%TLDR, needs to be formulated better:
%\begin{itemize}
%    \item Machine learning benefits from adding prior knowledge to the problem.
%    \item Machine learning frameworks under the hood are toolboxes for automatic differentiation and gradient-based optimization of functions.
%    \item You can directly write a physical model in such a framework and use optimizer on models parameters.
%    \item We do so on light propagation model through a adaptive optics setup. We model a confocal microscope with scatterer in the path (phase object) with unknown surface. This
%\end{itemize}

%3. gap in research, problem we will address

% \js{is that what you meant?}
%\textcolor{blue}{Additionally using a pretrained model a question always arises whether the model input is valid } 
 
\begin{figure}
    \centering
    \includegraphics[width=0.85\textwidth,trim={0 2cm 0 0},clip]{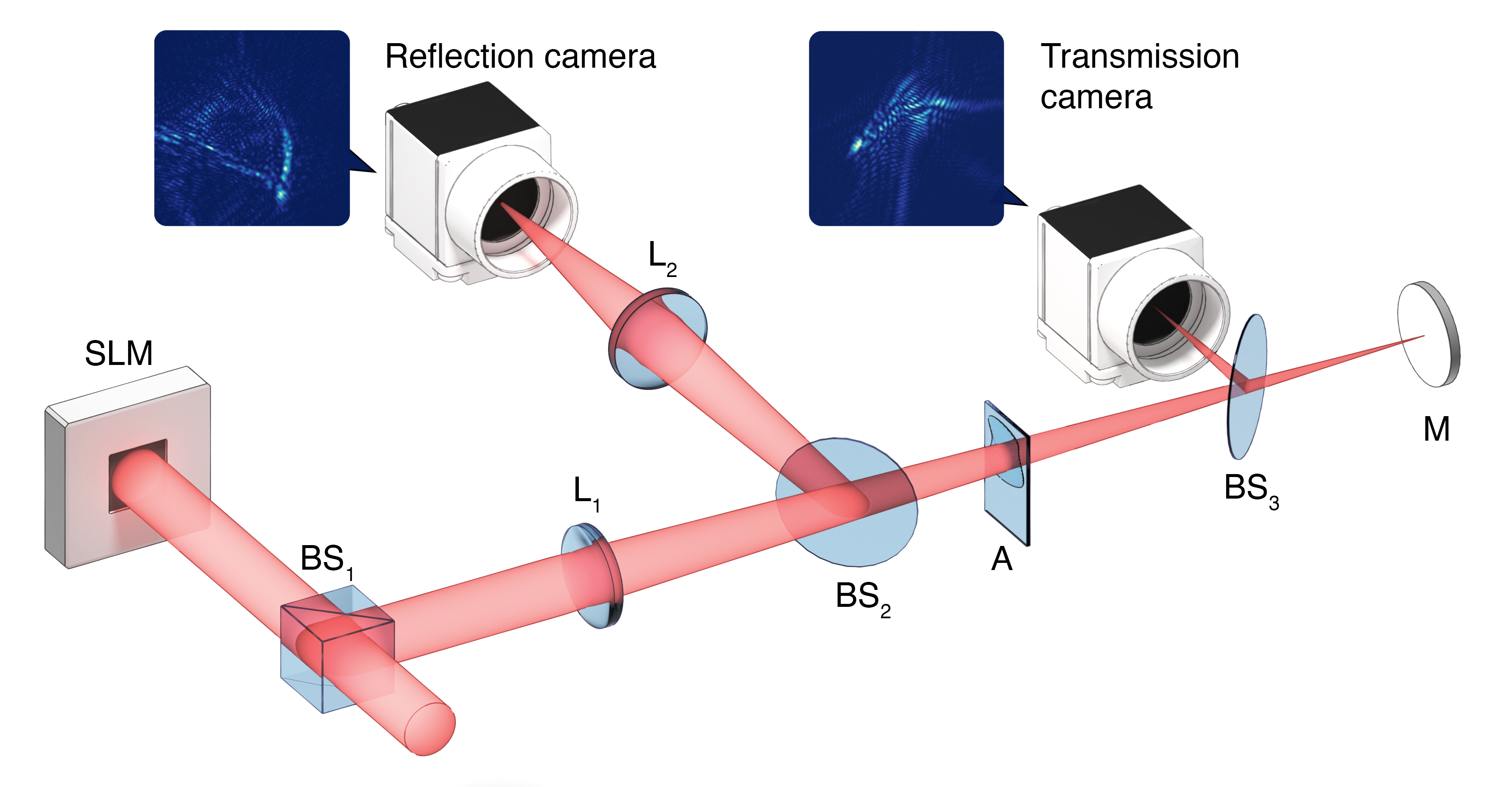}
    \caption{Schematic of experimental setup. Light reflected off a spatial light modulator (SLM) passes through an aberration (A) and is focused onto a camera (transmission camera, shown with illustration of imaged light distribution). For experiments in an epidetection configuration, light reflected off the mirror M at the sample plane is additionally recorded with a second camera (reflection camera). BS = beam splitter, L = lens (see main text and Methods for details).}
    %\nolinebreak
    \label{fig:setup}
\end{figure}

%4. describe the paper
Here, we employ this model optimization strategy for adaptive optics: we describe light propagation through the optical system, including unknown aberrations represented as parameters, with a differentiable model (Fig.~\ref{fig:setup}). For matching the input-output relationship of the computational model to the experimental setup we record a number of output images resulting from corresponding input phase modulations and optimize model parameters using Tensorflow.  %(compared with the number of samples required for training a deep neural network for a similar task). 
We show that this allows extracting an accurate description of the introduced aberrating layer(s) as verified by focusing in transmission through a single layer, as well as in a reflection, through two layers. In the latter epidetection configuration only reflected light is used for optimization and transmission focusing.

%We further show that this approach canBy learning inverse models of aberrations,  be optimized be additionally including a deep image prior into the model. Such an untrained network model speeds up the optimization process and improves over multiple training runs. 

%%%%%%%%%%%%%%%%%%%%%%%%%%%%%%%%%%%%%%%%%%%%%%%%%%%%%%% 
%\section{Results}

%%%%%%%%%%%%%%%%%%%%%%%%%%%%%%%
%FIGURE LIST
%fig1: a) conceptual schematic b) experiment schematics
%fig2: a) optimization in transmission: 
%a) training example, matched patterns after training, 
%b) aberrated and corrected focus 
%Supplementary: correlation coefficient for all )
%fig 3) same in reflection
%a) training example, matched patterns after training, 
%b) aberrated and corrected focus 
%c) optimization with blank sample in transmission and blank sample in reflecion
%Supplementary: correlation coefficient for all )

\section*{Results}

The experimental setup is shown schematically in Fig.~\ref{fig:setup}. An expanded and collimated laser beam is reflected off a spatial light modulator (SLM)  with a beam splitter cube (BS$_1$) and a part of the beam is imaged onto a camera (transmission camera) over a beam splitter (BS$_3$). For experiments with reflected light, the remaining part of the beam is additionally sent to a mirror at the sample plane (same focal plane as the transmission camera) which serves as a proxy for a reflecting sample. Light reflected by the mirror is imaged onto a second camera (reflection camera) with a beam splitter (BS$_2$). Aberrations (a layer of nail polish on a microscope slide, see Methods) are introduced between beam splitters BS$_2$ and BS$_3$. The beam undergoes aberrations once to the transmission camera, and twice to the reflection camera.

%The model of the setup \ref{fig:setup} is therefore a function composed of several factors of wavefront propagation and phase objects (for a single forward pass):
A single transmission pass through the setup is described by the function (see Methods for details)
\begin{equation}\label{eq:simulation}
S(\phi_\mathrm{SLM}, \phi_\mathrm{aberration}) = \left|P_{f_1}(\exp\left[i\phi_{\mathrm{aberration}}\right]\exp\left[i\phi_{\mathrm{lens}}\right] P_{f_1}(U_{0}\exp\left[i\phi_\mathrm{SLM}\right]))\right|^2,
\end{equation}
where $P_{d}$ is a propagation operator over the distance $d$, $U_\mathrm{0}$ is the complex amplitude of the unmodulated beam at the SLM, $\phi_{\mathrm{lens}}$ is the phase representation of the lens $L_1$, and $f_1$ is its focal length; $\phi_\mathrm{SLM}$ is the (known) SLM phase modulation, and $\phi_{\mathrm{aberration}}$ is the (unknown) introduced aberration. 

For computational efficiency, the aberration is simulated at the same plane as lens $L_1$ (see Methods). %A similar expression results for the combined transmission and reflection pathway (see Methods).
Finding the unknown aberration which maximizes the similarity, measured with Pearson's correlation coefficient $r$, of the simulated camera images $S(\phi_\mathrm{SLM}, \phi_\mathrm{aberration})$ and experimentally recorded images $I$ was solved in Tensorflow using automatic differentiation and gradient-based optimization (see Methods):
\begin{equation}\label{eq:argmax}
\phi_{\mathrm{aberration}} = \operatorname*{arg\,max}_{\phi_{\mathrm{aberration}}} (r\left[S(\phi_{\mathrm{SLM}}, \phi_{\mathrm{aberration}}), I\right]).
\end{equation}

To further refine the focus after a first optimization step, a second step was performed with a new set of modulations and corresponding images. % by applying the found correction to the SLM and repeating the procedure to find the residual uncompensated aberration. 
In this second step, the correction obtained in the first optimization step was added to all modulations displayed, $\phi_{\mathrm{SLM}}+\phi_{\mathrm{correction_1}}$. The final correction was the sum of the first and second step correction $\phi_{\mathrm{correction_1}}+\phi_{\mathrm{correction_2}}$. We used 180 modulations for transmission and 540 for reflection experiments in each of the two iteration steps.
%Optimization was performed in two steps, each with an independent set of modulations fist approximation served as the starting point of a second optimization run, again with xxx samples. The same two-step technique was used in reflection, but with XXX sample in each step.

\begin{figure}
    \centering
    \includegraphics[width=\textwidth]{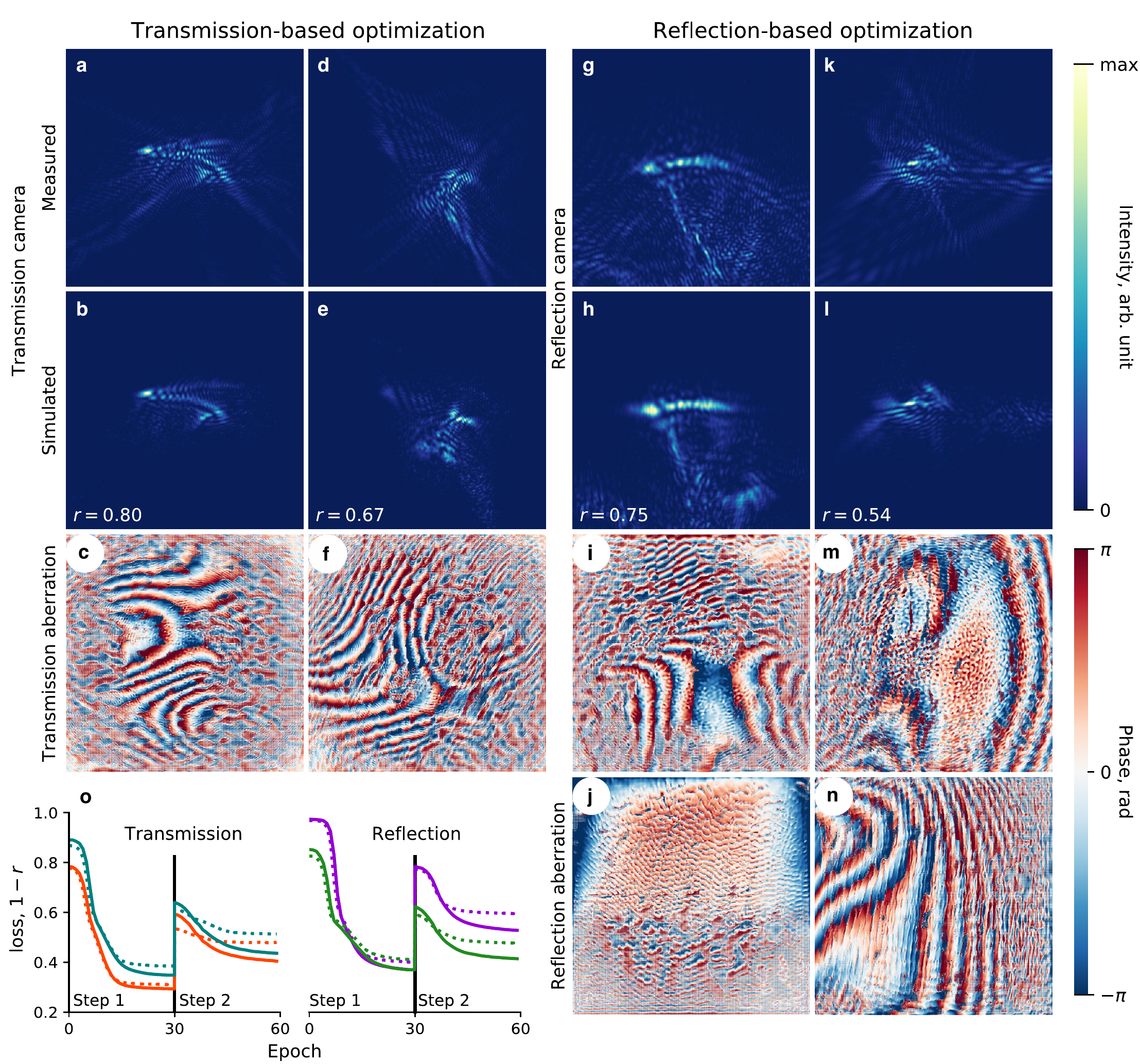}
    \caption{Model matching and optimization: \textbf{a-f} Transmission-based. \textbf{a, d} Two examples of measured, aberrated transmission light distributions and \textbf{b, e} matching simulated light distributions after first optimization step. \textbf{c, f} Corresponding phase profile (obtained at lens plane, after first optimization step).
    \textbf{g-n} Reflection-based.
    \textbf{g, k} Two examples of measured, aberrated, reflection light distributions and \textbf{h, l} matching simulated light distribution after first optimization step. \textbf{i, m} Corresponding transmission and \textbf{j, n} reflection phase results after first optimization step at the planes of lenses $L_1$ and $L_2$, respectively. \textbf{o} Transmission, orange: loss function for training of first example (a-c), and validation (dotted). Vertical line separates first and second optimization step, see Methods. Transmission, teal: same as orange for second example (d-f). Reflection, green: same for reflection experiment (g-j). Reflection, violet: same for second example in reflection experiment (k-n). $r$ in b, e, h, l is Pearson's correlation coefficient (with corresponding image in in a, d, g, k), field of view is 1766 $\mu$m by 1766 $\mu$m.}
    \label{fig:experiment_correlations}
\end{figure}

\begin{figure}
    \centering
    \includegraphics[width=\textwidth]{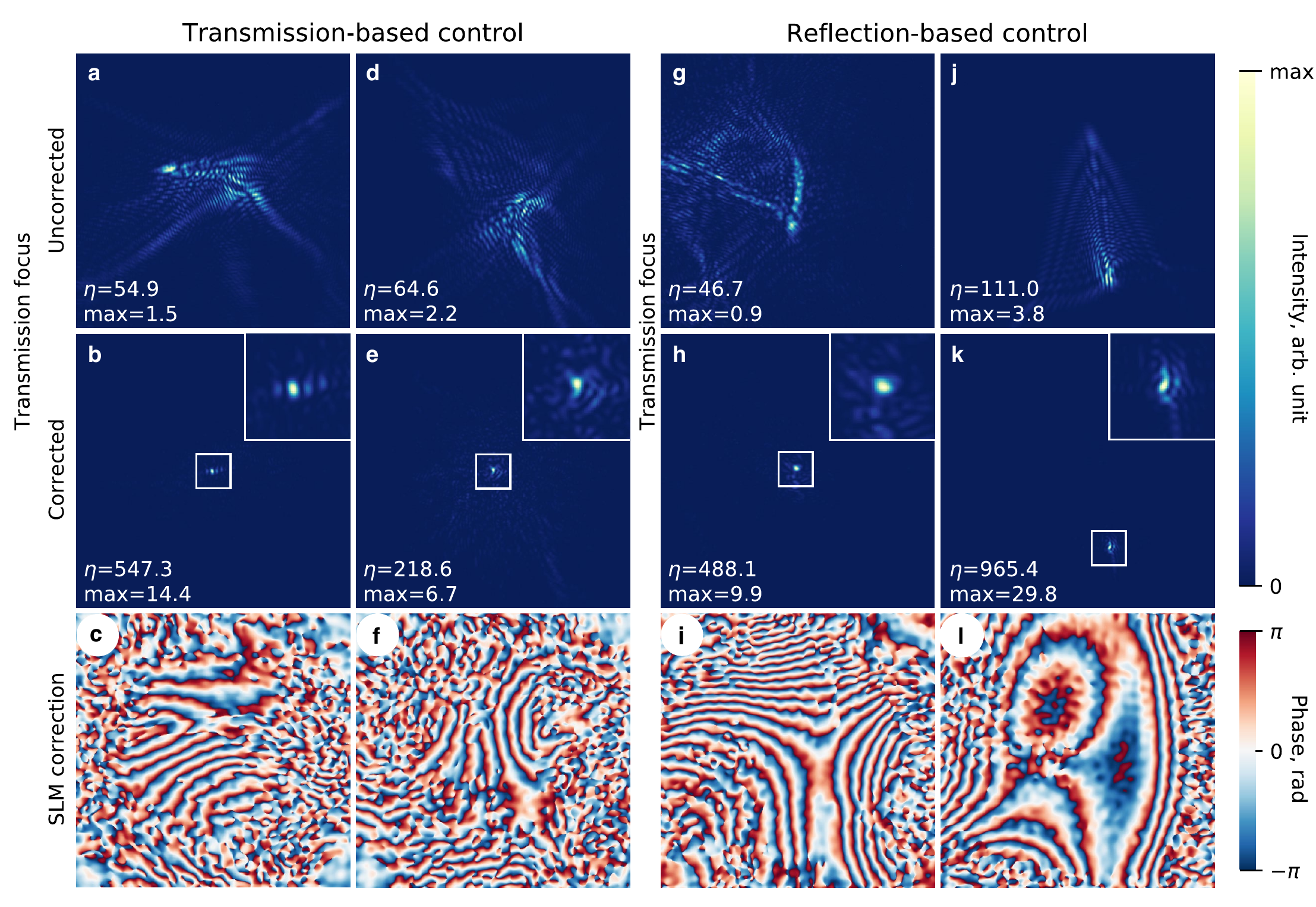}
    \caption{\textbf{a-f} Focusing in transmission. \textbf{a, d} Two examples of aberrations recorded in transmission. \textbf{b, e} Focus after correction with blow-up of focal spot (white frame and inset). \textbf{c, f} Wavefront correction at SLM after two-step optimization.
    \textbf{g-l} Focusing in transmission using reflected light. \textbf{g, j} Two examples of aberrations recorded in transmission. \textbf{h, k} Transmission focus after correction only using reflected light, with blow-up of focal spot (white frame and inset). \textbf{i, l} Transmission wavefront correction at SLM after two-step optimization recovered from only reflected light measurements. In each subfigure, max indicates maximum of colorbar, $\eta$ is enhancement (see Methods), field of view is 1766 $\mu$m by 1766 $\mu$m.
    }
    \label{fig:experiment_foci}
\end{figure}

As seen for two representative examples in Fig.~\ref{fig:experiment_correlations} a, b and d, e, respectively,  optimization (which results in the corresponding phase profile in Fig.~\ref{fig:experiment_correlations} c, f) leads to closely matching (correlation coefficient $r$ is indicated in Fig.~\ref{fig:experiment_correlations} b, e) measured and predicted light distributions at the sample (transmission camera) after applying the correction at the SLM. The similarity is quantified with the loss function $1-r$ in Fig.~\ref{fig:experiment_correlations} o.
To verify the achieved correction, the optimized aberration found at the plane of lens $L_1$ was propagated back to the SLM (see Methods) and the corresponding correction (Fig.~\ref{fig:experiment_foci} c and f) was displayed. This led to a focus at the sample or camera plane as shown with two representative examples in Fig.~\ref{fig:experiment_foci} b, e (Fig.~\ref{fig:experiment_foci} a, d shows the focus before correction) together with an increase in enhancement by a factor of 10 and 3.4, respectively (see Methods for definition of enhancement and details). (Note the difference in color scale between different images, normalized to maximum (max) values indicated in the subfigures.)

%\begin{figure}
%    \centering
%    \includegraphics[width=\textwidth]{figures/fig_reflection_edit.pdf}
%    \caption{Focusing in reflection using transmitted light. \textbf{a, c} Two examples of aberrations recorded in transmission and \textbf{e, g} corresponding flat SLM modulation. \textbf{b, d} Transmission focus after correction with blow-up of focal spot (white frame). \js{size of field of view}. \textbf{f, g} Displayed transmission wavefront correction after two-step optimization recovered from reflected light measurements.}
%    \label{fig:experiment_reflection}
%\end{figure}

In an epidetection configuration, as typical for imaging in biological samples, only reflected light can be used for finding a correction. Reflected light however accumulates a first aberration encountered in the excitation pass and a (generally different) second aberration in the reflection pass \cite{yoon2020deep, vishniakou2020wavefront}. These need to be disentangled for example to recover a transmission pass correction required for generating a focus inside a sample. %In the current frameworks, this corresponds to propagation through two scattering layers, since generally the wavefront can travel through different parts of the sactterer on the transmission and reflection passes \cite{vishniakou2020wavefront}.
%, a situation similar to one encountered in mutliconjugate adaptive optics (refs). 
For focusing in transmission using only reflected light therefore the function $S$ (equation \ref{eq:simulation}) is extended (see Methods) to include the reflection pass from the mirror at the sample plane through the aberration to the second camera (reflection camera), now including an aberration in the transmission as well as in the reflection pass. This model is fitted to match observed reflected light distributions by optimizing at the same time two independent aberrations. Optimization is performed as before (see Methods). 

Two representative examples of predicted and measured light distributions at the sample plane (transmission camera) are shown in Fig.~\ref{fig:experiment_correlations} g, h and k, l, respectively, and the loss function quantifying the similarity is shown in Fig.~\ref{fig:experiment_correlations} o (correlation coefficient $r$ between predicted and measured distributions is indicated in Fig.~\ref{fig:experiment_correlations} h, l). The corresponding transmission and reflection phase aberrations at the plane of lens $L_1$ and $L_2$ are shown in Fig.~\ref{fig:experiment_correlations} i, j and m, n, respectively. 
To verify the correction, we generated a focus at the sample plane by displaying the corresponding transmission correction on the SLM (see Methods). Fig.~\ref{fig:experiment_foci} shows two representative examples (g-i and j-l) of aberrated focus, corrected focus, and corresponding correction (resulting in an increase in enhancement by a factor of 10.4 and 8.7, respectively, see Methods). In reflection-based transmission control, the obtained focus was not necessarily centered in the field of view (as for example seen in Fig.~\ref{fig:experiment_foci} j, k), due to tilt introduced by the sample that was not corrected. Importantly, in reflection-based transmission control experiments, focusing in transmission is achieved only using reflected light, compatible with an epidetection configuration.

For reflection-based transmission control, aberrations in two different focal planes are independently computed at the same time. This is similar to multiconjugate adaptive optics, where typically however an additional SLM is used to correct for a second focal plane \cite{rigaut2018multiconjugate, kam2007modelling, thaung2009dual}. %Using two SLMs to independently correct the transmission and reflection aberration would be expected to further improve the corrections. %Techniques to control phase images in multiple planes are also of interest in holography \cite{tsang2016review, sinclair2004interactive, makowski2007iterative, ying2011new}, and could be achieved similarly using optimization by replacing the unknown aberrating layer with a desired target phase mask.
%shoudl cite: Laura Waller: 3D computer-generated holography by non-convex optimization
Additionally, while we generated single focal spots, arbitrary other focal distributions could be generated as well (for example for applications in optogenetics).

\section*{Methods}
%THIS SECTION IS NOT INCLUDED IN THE 1500 WORD COUNT

\subsection*{Experimental setup and data acquisition}

The laser was from Toptica (iBeam smart, 640 nm), the spatial light modulator from Meadowlark (SLM, ODP512-1064-P8), cameras were from Basler (acA2040-55um). All optical parts were from Thorlabs: $BS_1$ in Fig.~\ref{fig:setup} was BS016, $BS_2$ and $BS_3$ were EBS2. Lenses were visible achromats, $L_1$ with focal length 300 mm (AC254-150-A) and $L_2$ with 150 mm (AC254-150-A).
Data were collected by placing an aberrating sample in the optical path between BS2 and BS3 (see Fig.~\ref{fig:setup}), displaying random SLM phase modulations, and recording the resulting $512\times512$ images with the transmission camera, and additionally with the reflection camera for reflection experiments. Random SLM phase modulations were generated by summing the first 78 Zernike modes with random coefficients drawn from a normal distribution with standard deviation $\pi$ and displayed at a resolution of $512\times512$ pixels on the SLM.  
%with a beam splitter cube ($BS_1$ in Fig.~1, \js{part number}) and imaged onto a camera (camera 1 in Fig.~ 1, Basler acA2040-55um) as well as a reflecting mirror (M) at the same focal plane with a second beam splitter ($BS_2$, , Thorlabs). Light reflected by the mirror (which serves as a proxy for light reflected by a scattering sample) travels back through the beam splitter and the mirror is imaged onto a second camera with another beam splitter ($BS_3$, EBS2, Thorlabs) and a lens ($f_2 = 2 \cdot f_1$, XXX, Thorlabs). An aberrating layer (a thin layer of transparent nail polish distributed on a a micorscope slide) is inserted between the two beam splitters.

The light intensity in transmission and reflection can vary by several orders of magnitude, exceeding the dynamic range of the cameras. Therefore, in order to capture the full range of intensities, multiple frames with different exposure times were recorded (each frame with a 12-bit per pixel resolution). In transmission, each frame was recorded with exposures of 60, 120, and 250 ms, respectively, and the resulting image was the sum of the recorded frames weighted by the inverse of the exposure time. Saturated pixels as well as pixels below the noise threshold were discarded. For the reflection camera, images were taken with exposures of 60, 120, 250, 500, and 1000 ms, respectively. Additionally, transmission light intensity was reduced with a neutral density filter wheel (NDM2/M, Thorlabs). 

\subsection*{Computational model}
Light travelling through the setup is modeled as a complex amplitude $U(x, y, z)$, initialized with $U_0 = U(x, y, 0)$, and propagating through a sequence of planar phase objects and intermittent free space along the optical axis ($z$-axis; x, y, z are spatial coordinates). A wavefront $U(x, y, d)$ interacting with a phase object $\phi(x, y, d)$ at plane $d$ is described as a multiplication
%\js{something wrong with this formula? U factor is the same on both sides}
\begin{equation}\label{eq:phase_interaction}
%U(x, y, d)=U(x, y, d)\exp\left[{i\phi(x,y,d)}\right].
U(x, y, d)\exp\left[{i\phi(x,y,d)}\right].
\end{equation}
Free space propagation of the wavefront over a distance $d$ is calculated using the angular spectrum method  with the following operator \cite{goodman2005introduction}:
\begin{equation}\begin{aligned}\label{eq:propagation_operator}
U(x, y, z+d) &= P_d(U(x, y, z)) = \iint A(f_X, f_Y; z)\,\mathrm{circ}\left(\sqrt{(\lambda f_X)^2+(\lambda f_Y)^2}\right)\\&\times H\exp\left[i2\pi(f_Xx+f_Yy)\right] \,\mathrm{d}f_X\,\mathrm{d}f_Y.
\end{aligned}\end{equation}
Here, $A(f_X, f_Y; z)$ is the Fourier transform of $U(x, y, z)$, $f_X$, $f_Y$ are spatial frequencies, the circ function is 1 inside the circle with the radius in the argument and 0 outside \cite{goodman2005introduction},
%$\left(\sqrt{(\lambda f_X)^2+(\lambda f_Y)^2}\right) < 1$
%$\mathrm{circ}\left(\sqrt{a^2+b^2}\right) = 1$ for $\left(\sqrt{a^2+b^2}\right) < 1$ , 
%$= 0.5$ for $\left(\sqrt{a^2+b^2}\right) = 1$ , 
%and $= 0$ for $\left(\sqrt{a^2+b^2}\right) > 1$. 
%the low pass filtering function
and $H(f_X, f_Y) = \exp\left[i2\pi\frac{d}{\lambda}\sqrt{1-(\lambda f_X)^2-(\lambda f_Y)^2}\right]$ is the optical transfer function. The intensity measured by the camera is given by 
\begin{equation}\label{eq:intensity}
I(x, y, z) = \left|U(x, y, z)\right|^2.
\end{equation}

\subsection*{Model optimization}
We used a Python library for diffractive optics \cite{diffractio} to calculate the known factors of (\ref{eq:simulation}). By providing the focal lengths and setup dimensions, discretized versions of the optical transfer functions for the propagation operators and the phase representation of the lenses were determined. The resulting function which relates displayed, known SLM phase modulations and unknown sample aberrations to camera images, was then transferred to Tensorflow.   %allowing to take advantage of automatic differentiation and gradient-based optimization 
The position of the scatterer, as seen in expression (\ref{eq:simulation}), was simulated at the plane of lens $L_1$. This saves computations and memory, since each intermediate plane requires additional wavefront propagation calculations. Similarly, for computational efficiency, a single lens $L_2$ with focal length $f_2=\frac{f_1}{2}$ was used to focus reflected light onto the reflection camera.
While the parameters of the optical model for transmission and reflection were adjusted manually to match the setup, they can equally be tuned using the optimization approach described below, for example to obtain a systems correction.

The model of light propagation in the setup, equation (\ref{eq:simulation}), was incorporated in the loss function according to equation (\ref{eq:argmax}):
\begin{equation}\label{eq:loss}
    \mathrm{loss} = 1-r\left[S(\phi_{\mathrm{SLM}}, \phi_{\mathrm{aberration}}), I\right],
\end{equation}
where $\phi_{\mathrm{SLM}}$ and $\phi_{\mathrm{aberration}}$, are the phase modulations at the SLM and due to the introduced aberration, respectively. All variables are $512\times512$ real-valued tensors, and $\phi_{\mathrm{aberration}}$ is the optimization variable.

Similar to the training of neural networks, we split the data into training and validation sets and used batches. We used Adam optimizer with learning rate 0.1 and batch size 30. Optimization with the loss function resulted in matching simulated and experimentally recorded images and yielded the  phase profile of the aberration in the setup. The quality of the solution was quantified with the correlation between modelled and recorded images in the validation part of the dataset and was used as the criterion for stopping the optimization. Convergence of the optimization process depended on the magnitude and spatial frequencies of the aberration and the number of samples. %The success of model fitting is described by the correlation coefficient $r$ in (\ref{eq:argmax}).
Typically a solution with $r>0.6$ was sufficient for focusing.

For experiments with reflected light, the simulation of the setup was extended to include the reflected light pass,
\begin{equation}\begin{aligned}\label{eq:simulation2}
S(\phi_\mathrm{SLM}, \phi_\mathrm{trans}, \phi_\mathrm{refl}) &=  \left|\right.P_{f_1}(\exp\left[i\phi_{\mathrm{lens_2}}\right]\exp\left[i\phi_{\mathrm{ref}}\right]\\
&\times P_{2\cdot f_1}(\exp\left[i\phi_{\mathrm{trans}}\right]\exp\left[i\phi_{\mathrm{lens}}\right] P_{f_1}(U_{0}\exp\left[i\phi_\mathrm{SLM}\right])))\left.\right|^2,
\end{aligned}\end{equation}
and the loss function (\ref{eq:loss}) is optimized with variables $\phi_{\mathrm{trans}}$ and $\phi_{\mathrm{ref}}$.

\subsection*{Evaluation}

After $\phi_{\mathrm{aberration}}$ is found through optimization at the lens plane, the corresponding correction at the SLM is found by propagating the conjugate phase of the aberration backwards to the SLM plane $\phi_{\mathrm{correction}}= \mathrm{arg}(P_{-f_1}(\exp\left[-i\phi_{\mathrm{aberration}}\right]))$. Additionally, we smooth the found correction with a low-pass spatial frequency filter: discrete Fourier transform is applied to $\mathrm{exp}\left[i\phi_{\mathrm{correction}}\right]$ and frequencies exceeding $0.1$ of the pattern resolution are discarded. When displayed on the SLM, $\phi_{\mathrm{correction}}$, this results in a compensation of the aberration.

Aberrations were introduced with a thin layer of transparent nail polish distributed on a microscope slide (inserted between $BS_2$ and $BS_3$). Two different aberrating samples were used for transmission and reflection experiments. Generally, the strength of aberrations varies depending on sample positioning. Optimization parameters (such as number of phase modulations or learning rate) %can be optimized for each aberration. Here, 
were only adjusted once for transmission experiments, and once for reflection experiments. As a simple measure for quantifying the shape of the uncorrected light distributions, we use its  maximum extension as measured by the length of the first principle component of the pixels above a 30 \% intensity threshold, $\sigma$. To quantify the change in the distribution before and after correction we compared the uncorrected and corrected distribution, $\sigma_{\mathrm{rel}} = \sigma_\mathrm{u}/\sigma_\mathrm{c}$.  To additionally quantify the quality of the aberration correction, we also used an enhancement metric defined as ratio of maximum intensity to mean intensity in the frame, $\eta=\mathrm{max}(I)/\mathrm{mean}(I)$, comparing it before and after correction $\eta_\mathrm{rel}=\eta_\mathrm{c}/\eta_\mathrm{u}$. The distribution of these values ($\eta_\mathrm{rel}, \sigma_\mathrm{rel}$) for a series of 7 transmission experiments was: (10.0, 25.3), (1.7, 6.5), (3.4, 12.1), (2.6, 10.4), (0.8, 1.2), (16.2, 25.4), (3.8, 9.8), $\langle\eta_\mathrm{rel}\rangle=5.5\pm5.2$, $\langle\sigma_\mathrm{rel}\rangle=13.0\pm8.5$; and in a series of 7 reflection-based transmission control experiments was: (10.5, 17.6), (10.4, 32.6), (11.6, 11.1), (7.2, 13.0), (8.7, 4.3), (4.4, 12.3), (1.9, 2.6), $\langle\eta_\mathrm{rel}\rangle=7.8\pm3.3$, $\langle\sigma_\mathrm{rel}\rangle=13.4\pm9.2$.

\section*{Discussion}
Differentiable model-based approaches for image reconstruction have been introduced in several domains of imaging \cite{ongie2020deep}, for example in ptychography \cite{kellman2019data}. Instead of directly optimizing model parameters, an additional deep neural network (a deep image prior \cite{ulyanov2018deep, heckel2019denoising}) has also been introduced, for example for phase imaging \cite{wang2020phase, bostan2020deep} or ptychography \cite{zhou2020diffraction}.%, corresponding to additional regularization of the solution.
Even without such additional regularization, the optimization converged reliably to smooth phase patterns (Fig.~\ref{fig:experiment_correlations} c, f, i, m, j, n and Fig.~\ref{fig:experiment_foci} c, f, i, l), but for example a DIP could also be combined with the introduced method to further reduce the number of samples used for optimization.

%3. achievements/contributions, refining implications
The number of required samples depends on the magnitude and spatial frequencies of the aberrations, requiring more samples with stronger aberrations. This can be compared to the training of deep neural networks, where the number of required samples for model training similarly increases with increasing aberrations.
%convergence of training, a requirement of obtaining a suitable model, becomes typically more challenging with increasing variety and order of scattering samples. 
%Model optimization in transmission was achieved in two steps, with 180 samples in each step. In reflection, again two optimization steps were employed, each with 540 samples.
Compared to deep neural networks, including a physical model of the imaging process allows finding aberration corrections with a small number of samples (albeit only for a single field of view at a time). Different from neural networks which are trained on a predetermined distribution of aberrations (for example based on Zernike polynomials), optimization is achieved independently in each pixel without prior assumptions about aberrations.

%4. Limitations, current and future work, applications
Similar to other techniques that require multiple measurements for finding a correction \cite{yoon2020deep, gonsalves2014perspectives}, a limitation of the presented approach for dynamic samples is the time it takes to find a correction. The optimization time, several minutes on a single GPU, could be reduced by using multiple GPUs. Generally, the gap between optimization and control (corresponding to rapidly changing corrections in response to aberrations) is expected to narrow with increasing computational power \cite{brunton2020machine}. %Alternatively, while we here used two steps of optimization to find a correction, more steps with fewer samples would lead to faster optimization convergence in a single step and could for example allow tracking slowly changing aberrations.

Thanks to advanced computational frameworks \cite{diffractio, tensorflow2015-whitepaper}, the introduced model-based optimization can easily be combined with any optical setup equipped with a spatial light modulator and a camera without requiring additional hardware such as wavefront sensors or interferometers. 
For example, the described technique could be combined with imaging through scattering materials in a microscope with a high numerical aperture objective in an epidetection configuration \cite{vishniakou2020wavefront}. In summary, we expect that the developed method will be useful in many situation that can benefit from correcting aberrations through single and multiple layers.% without being limited to a predetermined model of aberrations.

%Could mention that mixed methods, between pretrained NN and optimization approach could be developed? For example, use NN result for initialization or faster convergence?

%\js{for discussion: the position of the scatterer could also be optimized}

%%%%%%%%%%%%%%%%%%%%%%%%%% Tail %%%%%%%%%%%%%%%%%%%%%%%%%%%%%%

\section*{Funding}
This work was supported by the Max Planck Society and the research center caesar. 
 
%\section*{Acknowledgments}

\section*{Disclosures}
The authors declare no conflicts of interest.

%%%%%%%%%%%%%%%%%%%%%%% References %%%%%%%%%%%%%%%%%%%%%%%%%
\bibliographystyle{unsrt}
\bibliography{submission}

\end{document}